\documentclass[twocolumn,showpacs,preprintnumbers,amsmath,amssymb]{revtex4}

\usepackage{graphicx}
\usepackage{bm}

\newcommand{\beq}{\begin{equation}}
\newcommand{\eeq}{\end{equation}}
\newcommand{\bea}{\begin{eqnarray}}
\newcommand{\eea}{\end{eqnarray}}

\newcommand{\eg}{e.g. }

\let\a=\alpha \let\b=\beta  \let\g=\gamma   
       \let\l=\lambda
             \let\p=\pi    
    \let\f=\varphi

    \let\Si=\Sigma

\let\io=\infty

\begin{document}


\title{Glassy behavior of light}

\author{L. Angelani$^1$, C. Conti$^{2,3}$, G. Ruocco$^{3,4}$, F. Zamponi$^{3,4}$}

\affiliation{
$^1$Research center SMC INFM-CNR, c/o Universit\`a
di Roma ``La Sapienza,'' I-00185, Roma, Italy \\
$^2$Centro studi e ricerche ``Enrico Fermi,'' Via Panisperna 89/A,
I-00184, Roma, Italy   \\
$^3$Research center Soft INFM-CNR, c/o Universit\`a di
Roma ``La Sapienza,'' I-00185, Roma, Italy \\
$^4$Dipartimento di Fisica, Universit\`a di Roma ``La Sapienza,''
I-00185, Roma, Italy }


\begin{abstract}
We study the nonlinear dynamics of a multi-mode random laser using
the methods of statistical physics of disordered systems. 
A replica-symmetry breaking phase transition is
predicted as a function of the pump intensity. We thus show
that light propagating in a random non-linear medium displays
glassy behavior, i.e. the photon gas has a multitude of metastable
states and a non vanishing complexity, corresponding to
mode-locking processes in random lasers. The present work reveals
the existence of new physical phenomena, and demonstrates how
nonlinear optics and random lasers can be a benchmark for the
modern theory of complex systems and glasses.
\end{abstract}


\maketitle


The first marriage between statistical mechanics and lasers
dates back to their early development
\cite{HakenBook,RiskenBook}. Since the seventies,  many
authors have outlined that the threshold for lasing can be
interpreted as a thermodynamic phase transition; and these ideas
spread out in the field of photonics, later embracing also nonlinear
optics (for a review see e.g.  \cite{Arecchi99}).
In recent articles,  \cite{Vodonos04, Gordon03b, Gordon03, Gordon03c,Weill05} 
the statistical properties of laser light in {\it homogeneous
cavities} has been studied taking into account nonlinear phenomena, like gain saturation
and intensity dependent refractive index. This nonlinearity
gives rise to an interaction among the oscillation modes,
which in turn produces new interesting effects. Specifically, by
mapping the dynamics in an {\it ordered} Hamiltonian
problem, the authors of Ref.~\cite{Weill05} predicted a critical
behavior (a phase transition) of the laser
mode-locking process.

In the latter example, the statistical mechanics of {\it ordered}
systems is applied to study light propagation in amplifying
homogeneous non-linear materials.
In recent years, the locally inhomogeneous character of matter, and
in particular the disordered nature of these inhomogeneities, are
becoming more and more important. Specifically, relevant attention
has been dedicated to light amplification in random media and
random lasers
\cite{Lethokov68,Lawandy94,Wiersma95,Cao99,Wiersma01,Florescu04}. This is a
fascinating topic that bridges various fields like light localization and
diffusion, thermodynamics,  nonlinear physics and quantum 
optics, and it has
relevant fundamental and applicative perspectives, as in bio-physics \cite{Polson04}. Methods of statistical
mechanics have not yet been applied to the propagation of light in
non-linear {\it disordered} active media, but 
the rich behavior observed in glasses 
(aging, memory effects, etc.) can be foreseen ~\cite{MPVBook,Cu02}.

In this Letter, we analytically study the statistical properties
of the modes of a random optical cavity. Various physical settings
are embraced by this problem: for example a micro-structured
cavity filled by an active soft-material, like doped liquid crystals,
cavity-less random lasers, or even a standard laser system with a
disordered amplifying medium. In all these cases, the disorder 
varies on time scales much longer than the optical
fields (it is ``quenched'' \cite{MPVBook}) and, for any
realization of the system, the supported electromagnetic modes
interact because of the nonlinearity of
the resonant medium.

We first show that light propagation is described by an Hamiltonian
with the pumping rate acting as inverse
temperature. The mode amplitudes are taken as slowly varying and their phases
play the role of ``spins'' in an Hamiltonian, which turns out to be
a generalization of the XY model \cite{KT72} and of the so called
$k$-trigonometric model \cite{Ktrig}. The thermodynamics of the
model is studied using the replica trick \cite{MPVBook}, and a one
step replica symmetry breaking transition is found \cite{CC05}.
Our results show that when the average energy into each mode
increases (i.e. the ``temperature'' is decreased), the system
undergoes a glass transition, meaning that its dynamics
slows down and an exponentially large number of states appears,
with a non vanishing ``complexity'' \cite{CC05}. They
correspond to ``mode-locked''
states of a random laser. This treatment somehow generalizes to disordered systems  
early works on multi-mode lasers
(see e.g. \cite{Lamb64, Ducuing64, Bryan73}),
and can be also applied to other problems involving
multi-mode interactions in a nonlinear medium.
By providing an analytically treatable statistical model, not
only we unveil the complex behavior (in the meaning of modern
glassy physics) of light in active disordered media, but
we also demonstrate the possibility of using them for
testing the replica symmetry breaking transitions.
When studying the glass transition of {\it atomic} or {\it
molecular} systems, due to the predicted kinetic arrest, the
interesting time scale becomes extremely long with respect to that 
experimentally accessible. Due to the intrinsically fast
{\it photon} dynamics, it is reasonable
to expect that the system can be equilibrated much closer to the transition,
 and this should provide new and interesting
experimental data on kinetically arrested ``photon glasses''.


We consider a dielectric resonator
described by a refractive index profile $n({\bf r})$.
The time dependence of this quantity is of no interest here,
taking place on a time scale much slower than that of the photon
propagation.
Our approach follows the standard coupled mode-theory in the time
domain \cite{HausBook}.
The Maxwell's equations in the presence of a nonlinear polarization
$\mathbf{P}_{NL}$ are:
\begin{equation} \label{Max2}
\begin{array}{l}
\nabla\times\mathbf{\mathbf{H}}(\mathbf{r},t)=\varepsilon_0 n^2(r)
\partial_t\mathbf{E}(\mathbf{r},t)+
\partial_t \mathbf{P}_{NL}(\mathbf{r},t,\mathbf{E},\mathbf{H})\\
\nabla\times\mathbf{E}(\mathbf{r},t)=-\mu_0 \displaystyle
\partial_t \mathbf{H}(\mathbf{r},t) \ .
\end{array}
\end{equation}

The fields $\mathbf{E}(\mathbf{r},t)$ and
$\mathbf{H}(\mathbf{r},t)$ are conveniently decomposed in normal
modes corresponding to the solutions of the linear problem,
$\mathbf{P}_{NL}=0$, (frequencies $\omega_m$, eigenvectors
$\mathbf{ E}_m(\mathbf{r})$, $\mathbf{
H}_m(\mathbf{r})$, and ``amplitudes'' $a_m$). For later
convenience, they are cast in the form
\begin{equation} \label{eig1}
\begin{array}{l}
\mathbf{E}(\mathbf{r},t)=\sqrt{\omega_m} \; \; \Re [\sum_m a_m
{\mathbf{E}}_m(\mathbf{r})
\exp(-i\omega_m t)] \ ,\\
\mathbf{H}(\mathbf{r},t)=\sqrt{\omega_m} \; \; \Re [\sum_m a_m
{\mathbf{H}}_m(\mathbf{r}) \exp(-i\omega_m t)] \ ,
\end{array}
\end{equation}
where $\mathbf{
E}_m(\mathbf{r})$, $\mathbf{ H}_m(\mathbf{r})$
solve the eigenvalue problem
\begin{equation} \nonumber
\left(
\begin{array}{cc}
0 & i \nabla\times\\
-i \nabla\times & 0\\
\end{array}
\right) \! \! \left(
\begin{array}{c}
{\mathbf{E}}_m\\
{\mathbf{H}}_m
\end{array}
\right) \! = \! \omega_m \! \left(
\begin{array}{cc}
\epsilon_0 n^2 & 0 \\
0 & \mu_0\\
\end{array}
\right) \! \! \left(
\begin{array}{c}
{\mathbf{E}}_m\\
{\mathbf{H}}_m
\end{array}
\right)
\end{equation}
and are normalized such that
\begin{equation}
\frac{1}{4}\int_V \{ \varepsilon_0 n^2(r) \vert \mathbf{
E}_m(\mathbf{r})\vert^2+\mu_0 \vert\mathbf{
H}_m(\mathbf{r})\vert^2\} dV = 1 \ ,
\end{equation}
where $V$ is the standard normalization volume, over which
periodical boundary conditions are enforced.
With these notations, the total energy stored in the cavity is
$\mathcal{E} = \Sigma_m \mathcal{E}_m  = \Sigma_m \omega_m \vert a_m \vert^2$.

In the presence of nonlinearity, the amplitudes $a_m$ are time-dependent
and their evolution is described by coupled equations that can be derived
using standard perturbation techniques \cite{HausBook} and take the general form
\begin{equation}
\label{coupledmodes2} \frac{d
a_m(t)}{dt}=i\frac{\sqrt{\omega_m}}{4} \int_V \mathbf{
E}_m^*(\mathbf{r}) \cdot \mathbf{ P}_m(\mathbf{r}) \;  dV \ .
\end{equation}
The integral in Eq.~(\ref{coupledmodes2}) is extended to the
cavity volume where the nonlinear polarization is different from
zero. The quantity $\mathbf{ P}_m(\mathbf{r})$ is the
amplitude of the component of the nonlinear polarization
oscillating at $\omega_m$, $\mathbf{P}_{NL}=\Re[\sum_m
\mathbf{P}_m \exp(-i\omega_m t)]$ and its explicit expression as
function of the fields depends on the considered non-linearity. 
For isotropic media the leading cubic terms are
written, in our notation, as \cite{BoydBook}
\begin{equation} \nonumber
P_{m}^\alpha(\mathbf{r})= \!\!\!\!\!\!\!\!\!\!\!\!\!\!
\sum_{\omega_m=\omega_p+\omega_q-\omega_r}\!\!\!\!\!\!\!\!\!\!\!\!\!\!
K_{\alpha\beta\gamma\delta}
(\omega_m;\omega_p,\omega_q,-\omega_r,\mathbf{r}) E_p^\beta
 E_q^\gamma  E_r^\delta a_p a_q a_r^*.
\end{equation}
Here $K_{\alpha\beta\gamma\delta}$
$=\sqrt{\omega_p\omega_q \omega_r}$$\chi_{\alpha\beta\gamma\delta}$ and $\chi$ is
the third order response susceptibility tensor (explicit
expressions are known for example in the two level approximation
\cite{Bryan73,Bloembergen64,BoydBook}); the sum over Cartesian indeces is
implicit. The coupled mode theory
equations in a nonlinear cavity read hence as
\begin{equation}
\label{coupledmodes3} \frac{d a_m(t)}{dt}=-\frac{1}{2}
\Sigma_{pqr} g_{mpqr} a_p a_q a_r^*\text{,}
\end{equation}
where the sum is extended over all the modes and
\begin{equation}\label{gdef}
 g_{mpqr}\!=\!\!\frac{\sqrt{\omega_m}}{2i} \!\! \int_V \! \! \!\!
K_{\alpha\beta\gamma\delta}
(\omega_m;\omega_p,\omega_q,-\omega_r,\mathbf{r}) E_m^\alpha
E_p^\beta E_q^\gamma E_r^\delta dV\text{.}
\end{equation}
Under standard approximations \cite{Bryan73,BoydBook}, the tensor  $g$ is
a quantity symmetric with respect to the exchange of any couple of
indeces and can be taken as real-valued, neglecting processes like
self and cross-phase modulations between modes (S-XPM). 
These mechanisms
can be included in the model (see below), but will be here neglected for the sake of
simplicity. 
 Introducing the (real valued) function
\begin{equation}
H=\frac14 \Sigma_{spqr} g_{spqr} a_s a_p a_q^* a_r^* \ ,
\end{equation}
and taking into account radiation losses and material absorption
processes, represented by the coefficients $\alpha_m$ and light
amplification through $\gamma_m$ (see also \cite{HausBook}), the
resulting equation of motion for $a_m(t)$ is written as:
\begin{equation}\label{c4}
\frac{d a_m}{dt}=-\frac{\partial H}{\partial a_m^*}+(\g_m-
\a_m) a_m + \eta_m(t)  = - \frac{\partial \mathcal{H}}{
\partial a^*_m} + \eta_m(t) ,
\end{equation}
where $\mathcal{H}=\sum_m (\a_m-\g_m)|a_m|^2+H$, and
having introduced as usual the noise term  (see e.g. \cite{GardinerBook,RiskenBook}),
with $\langle \eta_p(t) \eta_q(t') \rangle=2 k_B T_{bath}
\delta_{pq}\delta(t-t')$, weighted by an effecting temperature $T_{bath}$, with $k_B$ the Boltzmann constant. 
The previous equation is a standard
Langevin model for a system of $N$ particles moving in $2N$
dimensions (represented by $\{\Re a_m,\Im a_m \}_{m=1..N}$)
\cite{RiskenBook,Gordon03} and its invariant measure is given by
$\exp(-\mathcal{H}/k_B T_{bath})$. Note that the latter is not affected by S-XPM terms,
as we will discuss in future publications and originally addressed in \cite{Gordon03}.

We consider a large number of modes in a small
frequency interval $\omega_m$$ \sim$$ \omega_0$ pumped and put into
oscillations. We can write $a_m(t)=A_m(t) \exp[i\varphi_m(t)]$,
which is useful as we expect $A_m$ to be slowly varying with
respect to $\varphi_m$, as discussed below. 
In previous works, with reference to standard multi-mode lasers, \cite{Ducuing64,Bryan73}
 the phase-dependent terms in (\ref{c4}), were always averaged out by assuming the
$\varphi_m$ rapidly varying, independent and uniformly distributed.
The resulting equations determine the oscillation energy $\mathcal{E}_m$ into each mode.
However, 
for an increasing number of modes, nonlinear beatings induce non-trivial light dynamics
which is mainly due to the rapidly varying phases, while the amplitudes 
can still be taken as slowly varying \cite{Ducuing64,HausBook,YarivBook}.

Summing up, the Hamiltonian depending on the relevant dynamic
variables, the phases $\varphi_m$, is written as
\begin{equation} \label{Ham}
{\cal H}(G,\varphi)={\cal H}_o + \Sigma_{spqr} G_{spqr}
\cos(\varphi_s +\varphi_p -\varphi_q -\varphi_r)
\end{equation}
where ${\cal H}_o=\sum_m (\alpha_m-\gamma_m)A_m^2$ is an
irrelevant constant term and $G_{spqr}$=$g_{spqr} A_s A_p A_q
A_r$. Hereafter we will consider these $G$ coefficients as
``quenched'' (due to the slow $t$ dependence of $A_m$), and the
relevant phase space is reduced to that spanned by $\varphi_m$.

If the cavity is realized by a random medium, as described above,
the coupling coefficients $G$ are random variables. Their values
depend on the mode profiles, the resonant frequencies, and on the
quenched values of the energies $\mathcal{E}_m=\omega_m A^2_m$ in each
mode, which vary with each realization of the cavity for a given
pumping rate. For these reasons we take $G_{spqr}$ as random
Gaussian variables. Due to the fact that the mode frequencies are
typically symmetrically distributed with respect to the transition
frequency of the amplifying atomic system (and correspondingly the
signs of the nonlinear susceptibility terms
largely varies, see e.g. \cite{BoydBook,Bloembergen64}) and their
spread is small, $\omega_m$$\sim$$\omega_o$, $G$ is taken with zero mean
value $\langle G\rangle=0$. The latter hypothesis can be removed
by a suitable, but not trivial, generalization of the treatment
below \cite{MPVBook}.

The $G$s roughly scale as $\langle A^2 \rangle^2  V^{-3/2}$,
as one can derive from Eq.~(\ref{gdef}) by using that $E^\a_m$
is $O(V^{-1/2})$ and $K_{\alpha\beta\gamma\delta}$ is a random variable integrated over $V$.
Recalling that $\omega_m \sim \omega_0$, $\omega_0 \langle A^2 \rangle$ measures the 
average energy per mode. 
By a simple rescaling, the invariant measure can be
written as $\exp[- \beta H(J,\varphi)]$, where
$J_{spqr}=G_{spqr}/ (g_0 \langle A^2 \rangle^2)$ has standard deviation
$\propto 1/V^{3/2} \propto 1/N^{3/2}$ and $g_0$ is a material-dependent constant. 
Note that this scaling of the $J$s
guarantees that the Hamiltonian is extensive \cite{MPVBook,CC05}. 
The parameter that controls the phase
transition is then $ \beta\equiv 1/ T \equiv \langle
A^2 \rangle^2 g_0 /k_B T_{bath}$. Thus, in what follows transitions obtained
as $ \beta$ is increased (i.e. the adimensional effective temperature $T$ is decreased), can be controlled by increasing the
amount of energy stored on average into each mode (i.e. the
pumping rate).

The calculation of the thermodynamics goes through the evaluation
of the partition function
\begin{equation}
Z(J) = \mbox{$\int$} d\varphi \, e^{- \beta {H}(J,\varphi)} \ ,
\end{equation}
and, more specifically, of the free energy $f( \beta)$ averaged
over the quenched disorder, that can be written, using the replica
 trick \cite{MPVBook}, as
\begin{equation}\label{freeenergy}
- \beta f( \beta)=\lim_{N \rightarrow \infty} \frac{1}{N}
\overline{\log Z(J)}=
 \lim_{N \rightarrow \infty} \lim_{n \rightarrow 0} \frac{\overline{Z^n(J)}-
1}{nN} \ . \nonumber
\end{equation}
By a standard saddle-point evaluation,
the replicated partition function is
$\overline{Z^n(J)} = \exp[-N n \min g(q)]$, where
the overlap matrix $q$ is defined by
$q_{ab}=N^{-1} \sum_j e^{i(\varphi_j^a-\varphi_j^b)}$
and one has to find
the overlap values $\overline q$
that satisfy the saddle-point equation.
The free energy is then given by
\begin{equation}
 \beta f = - \lim_{n \rightarrow 0} (nN)^{-1} \{e^{-Nn g(\overline
q)}-1 \} \sim \lim_{n \rightarrow 0} g(\overline q) =
 \b\phi(\overline q) \ , \nonumber
\end{equation}
where
\beq\label{f2min}
\begin{split}
& \b\phi(q)= -\frac{ \b^2}{4} +  \frac{3 \beta^2}{2} n^{-1} \Sigma_{a
> b} |q_{ab}|^4
 - n^{-1}\log Z(q) \ , \\
&Z(q) = \int d\varphi^a \exp \left[\Re \Sigma_{a>b} 2 \b^2
|q_{ab}|^{2} q_{ab}^* e^{i(\varphi^a- \varphi^b)} \right] \ .
\end{split}
\eeq This function is very similar to the one that describes the
Ising $p$-spin glass for $p=4$, see \eg \cite{CC05}. The
derivative with respect to $q_{ab}$ of $f(q)$ gives the saddle
point equation \beq q_{ab} = \langle \exp [i(\f^a-\f^b)] \rangle
\ , \eeq where the average is on the measure that defines $Z(q)$.
In the following, we will restrict ourselves to consider real
values of $q_{ab}$, without taking explicitly into account the
trivial rotational symmetry\cite{Binder86}.

The replica symmetric ({\sc rs}) solution corresponds to $q_{ab}\equiv q$ and
as usual it turns out that $q=0$ is the correct solution. The {\sc
rs} free energy is simply $f_{RS} = - \b/4$ as in the Ising
$p$-spin glass. In this regime the $\f_m$ are uniformly distributed in
$[0,2\p)$, independent and rapidly evolving, as originally considered in \cite{Bryan73,Ducuing64}.

In the one step replica symmetry breaking ({\sc 1rsb}) {\it ansatz}
one divides the matrix $q_{ab}$ in $n/m$ blocks of side $m$ \cite{MPVBook}.
 The
elements in the off-diagonal blocks are set to $0$ while in the
diagonal blocks {\sc rs} is assumed and $q_{ab}=q$.
In this case, Eq.~(\ref{f2min}) becomes:
\begin{eqnarray}
 \b\phi(m,q) =
 &-&( \b^2/4)\big[ 1 + 3 (1-m) q^4
 - 4 q^{3} \big]  \nonumber \\ &-& m^{-1}
 \log \mbox{$\int_0^\io$} dz \, z e^{-\frac{z^2}{2}} I_0( \b\l z)^m \ ,
\end{eqnarray}
where $I_n$ indicates the modified Bessel function of first kind
and $\l=\sqrt{2 q^3}$. The self-consistency
equation that defines $q$ is \beq\label{scons}
q\!=\frac{\mbox{$\int_0^\io$} dz \, z e^{-\frac{z^2}{2}} I^m_0( \b\l
z) \frac{I_1( \b\l z)^2}{I_0( \b\l z)^2}}{ \mbox{$\int_0^\io$} dz
\, z e^{-\frac{z^2}{2}} I^m_0( \b \l z)} \ . \eeq These
expressions are identical to the {\sc 1rsb} free energy for the
($p=4$) $p$-spin model, the only difference being the presence of
the Bessel function instead of the hyperbolic cosine in the last
integral. Starting from this expression one can repeat the
analysis of \cite{Montanari03} to derive the full phase space
structure of the model at the {\sc 1rsb} level.

\begin{figure}[t]
\includegraphics[width=250pt]{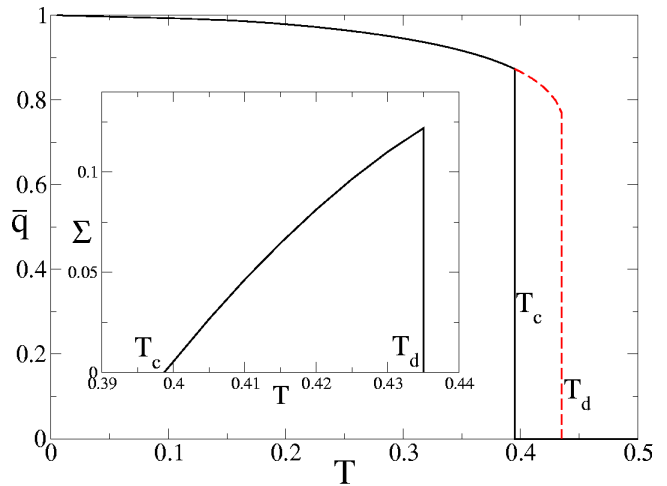} \vskip -0.5cm
\caption{The {\sc 1rsb} overlap $\bar q$ as a function of the reduced
temperature. The full line is the stable part of the curve, the
dashed line is the metastable part. Inset: the complexity $\Si( T)$
as a function of the temperature. It jumps discontinuously at
$ T_d$ where the metastable solution first appears and vanishes at
$ T_c$ where the glass transition takes place. } \label{figura}
\end{figure}

In Fig.~1 the solution $\bar q$ of Eq.~(\ref{scons}) for $m=1$ is
reported as a function of $ T$. At high temperature
$\bar q = 0$ and the {\sc rs} solution is recovered. On lowering
the temperature, a solution $\bar q \neq 0$ first appears at
$ T_d$ (dashed line). However, it becomes stable only
below the thermodynamic glass transition temperature $ T_c <  T_d$ (full line), as in
standard first-order transitions.  For $
T_c <  T <  T_d$ the phase space is
disconnected in an exponential number ${\cal N}( T) =
\exp N \Si( T)$ of states. The {\sc 1rsb} {\it
complexity} $\Si( T)$ is reported in the inset of
Fig.~1. At $ T= T_d$ a {\it dynamical
transition} is expected to take place \cite{CC05,Cu02}.

The stability of the {\sc 1rsb} solution can be studied (within
the assumption that $q_{ab}$ is real). It turns out that the {\sc
1rsb} solution is thermodynamically stable for all temperatures,
so in this case no Gardner transition is present
\cite{Ga85, Montanari03,Crisanti05}.

The presence of a dynamical phase transition at a given value of
the random laser pump intensity implies different interesting
physical phenomena. These could be experimentally investigated by studying
(for example via heterodyne experiments) the self correlation
function of a specific frequency ($\omega_m$) component of the
electric field in the cavity:
\begin{eqnarray}
C(t,\omega_m)= \omega_m A_m^2 \left \langle \exp \{ i [\varphi_m(t+\tau)-
\varphi_m(\tau)] \} \right
\rangle_{\tau} \ .
\end{eqnarray}
On approaching $ T_d$ from above (i.e. on increasing the
pump power), the dynamics of phase variable $\varphi_m(t)$ becomes
slower and slower and $C(t,\omega_m)$ is expected to decay towards
zero in longer and longer times. At the dynamical transition, the
dynamics of the $\varphi$'s becomes non-ergodic, they are no
longer able to explore the whole phase space and the function
$C(t,\omega_m)$ decays towards a plateau. In other words, the mode's
phases $\varphi_m(t)$ -beside from small oscillation- are locked
to some ``equilibrium'' values (``random mode locking''). Due to the
non-vanishing value of the complexity $\Sigma$ at $
T_d$, however, an exponentially large number of such equilibrium
positions exist, so giving rise to many different time structure
of the electric field in random lasers. On further increasing the
pump power, the complexity $\Sigma$ decreases 
and the ideal glassy state is reached at $
T$=$ T_c$. Below this value 
the number of equilibrium states is not exponential in $N$ ($\Sigma=0$). 
The equilibrium states below $T_d$ are difficult to reach because the needed time scale 
diverges \cite{Cu02}: interesting phenomena as aging, memory effects, and
history dependent responses are expected to take place for
$ T$$<$$ T_d$.

All these non equilibrium phenomena are theoretically predicted
and, to some extent experimentally
verified \cite{Cu02} in
material systems (structural glasses, spin glasses,..). The
absence of conclusive experiments is mainly due to the long time
needed to ``equilibrate'' real glasses below $ T_d$, a
time which is dictated by the elementary time of the {\it
atomic/molecular} dynamics in condensed matter (ps time scale).
{\it Photon} dynamics in cavities is at least three order of
magnitude faster. This observation drives us to propose the random
lasers in disordered media as a benchmark to test experimentally
the outcome of those theories, as the replica symmetry breaking. 

In conclusion, the multi-mode dynamics in a random laser cavity has been investigated by statistical
physics techniques, and a one step replica symmetry breaking phase
transition has been found. Our results emphasize two important points:
i) the light propagation in non-linear disordered media shows the
same complex behavior of the dynamics of glassy systems (aging,
memory, ...); ii) due to the faster photon dynamics with respect
to the atomic one, it is possible to use the random lasers as
systems for experimentally testing the replica symmetry breaking
transitions.

We are pleased to thank T. Castellani, L. Leuzzi, and F. Ricci-Tersenghi for useful discussions. 


\end{document}